\documentclass[12pt]{article}
\usepackage{graphicx}
\usepackage{epsf,amsmath,bbold,amsfonts,stmaryrd}

\usepackage[utf8]{inputenc}
\usepackage{mathrsfs}
\usepackage{appendix}
\usepackage{amssymb}
\usepackage{float}
\usepackage{color} 
\usepackage{cite}
\usepackage[colorlinks]{hyperref}
\hypersetup{pageanchor=false}
\usepackage{indentfirst}
\usepackage{url}
\usepackage{float}
\usepackage{caption}
\usepackage[numbers,square,comma,sort&compress,merge]{natbib}

\def\a{\alpha}

\def\B{\Box}

\def\d{\delta}
\def\D{\Delta}
\def\e{\varepsilon}

\def\eps{\varepsilon}
\def\f{\frac}

\def\G{\Gamma}

\def\l{\left}

\def\mc{\mathcal}

\def\m{\mu}
\def\n{\nu}

\def\nn{\nonumber}

\def\p{\partial}

\def\r{\right}
\def\s{\sigma}

\def\be{\begin{equation}}
\def\ee{\end{equation}}

\def\bea{\begin{eqnarray}}
\def\eea{\end{eqnarray}}

\def\ba{\begin{array}}
\def\ea{\end{array}}

\def\bc{\begin{center}}
\def\ec{\end{center}}

\def\bl{\begin{flushleft}}
\def\el{\end{flushleft}}

\def\br{\begin{flushright}}
\def\er{\end{flushright}}

\def\bi{\begin{itemize}}
\def\ei{\end{itemize}}

\def\bt{\begin{tabular}}
\def\et{\end{tabular}}

\begin{document}

\begin{titlepage}
\vspace{5cm}

\vspace{2cm}

\begin{center}
\bf \Large{Weyl vs. Conformal}

\end{center}

\begin{center}
{\textsc {Georgios K. Karananas, Alexander Monin}}
\end{center}

\begin{center}
{\it Institut de Th\'eorie des Ph\'enom\`enes Physiques, \\
\'Ecole Polytechnique F\'ed\'erale de Lausanne, \\ 
CH-1015, Lausanne, Switzerland}
\end{center}

\begin{center}
\texttt{\small georgios.karananas@epfl.ch} \\
\texttt{\small alexander.monin@epfl.ch} 
\end{center}

\vspace{2cm}

\begin{abstract}

In this note we show that given a conformally invariant theory in flat space-time,  it is not always possible to couple it to gravity in a Weyl invariant way.

\end{abstract}

\end{titlepage}

\section{Introduction}

The purpose of this note is to clarify the difference between the occasionally mixed notions of conformal and Weyl invariance. The conformal symmetry in a $n$-dimensional (not necessarily flat) space-time is defined as the group of coordinate transformations
\be
x' = F (x) \ ,
\ee
which leave the metric $g_{\m\n}$ invariant up to a conformal factor
\be
g _ {\m \n} (x) = \Omega (x') \, g _{\lambda \s}' (x') \, \f {\p F ^ \lambda} {\p x ^ \m} \f {\p F ^ \s} {\p x ^ \n} \ .
\label{metr_trans}
\ee
For the infinitesimal form of the transformations
\be
x'^{\m} = x ^ \m + f ^ \m \ ,
\ee
the relation~\eqref{metr_trans} leads to the conformal Killing equations
\be
\nabla _ \m f _ \s + \nabla _ \s f _ \m = \f {2} {n} g _ {\m \s} \nabla f \ ,
\label{conf_K}
\ee
where we used the shorthand notation $\nabla f = \nabla _ \m f ^ \m$, and we denoted with $\nabla$ the metric-compatible covariant derivative
\be
\nabla _ \m f _ \n = \p _ \m f _ \n - \G ^ \lambda _ {\m \n} f _ \lambda \ ,
\ee
with $\Gamma^\lambda_{\m\n}$ being the Christoffel symbols.

In this paper, we focus only on theories with scalars, leaving the investigation of fields with non-zero spin for elsewhere. The infinitesimal transformation of a scalar field with scaling dimension $\D$ under the full conformal group can be written in the following compact form
\be
\d _ c \phi = - \l ( f ^ \m \nabla _ \m \phi + \f {\D} {n} \nabla f \phi \r ).
\label{delta_c_phi}
\ee
A system is called conformally invariant if the variation of its action functional $S[g_{\m\n},\phi]$ under the full group of conformal transformations~\eqref{delta_c_phi} is zero, i.e.
\be
\d _ c S[g_ {\m \n}, \phi] = \int d ^ n x \, \f {\d S} {\d \phi} \, \d _ c \phi = 0 \ .
\label{conf_S}
\ee

Meanwhile, Weyl rescalings constitute another type of transformations, which are given by the simultaneous pointwise transformations of the metric and fields
\be
\begin{aligned}
\hat g _ {\m \n} (x) & = e ^ {2 \s (x)} g _ {\m \n} (x)~~~\text{and}~~~\hat \phi (x) & = e ^ {- \D \s} \phi (x) \ ,
\end{aligned}
\ee
with $\s$ being an arbitrary function. Writing the above expressions in their infinitesimal form as
\be
\d _ \s g _ {\m \n} = 2 \s g _ {\m \n} ~~~ \text {and} ~~~ \d _ \s \phi = - \D \s \phi \ ,
\label{Weyl_inf}
\ee
leads to the following condition for a theory to be Weyl invariant
\be
\d _\s S[g _ {\m \n}, \phi] = \int d ^ n x \, \s \l ( 2 \, \f {\d S} {\d g _ {\m \n}} g _ {\m \n} - \D_n \, \f {\d S} {\d \phi} \phi \r ) = 0 \ .
\label{conf_W}
\ee

Note that~\eqref{delta_c_phi} can be written as
\be
\d _ c \phi = \d _ d \phi + \d _ {\bar \s} \phi \ ,
\ee
where we denoted by $ \d _ {\bar \s}\phi$ the Weyl transformation corresponding to the  specific value of $\s = \bar \s \equiv \nabla f / n$, and $\d _ d \phi$ is the standard transformation of the scalar field under the general coordinate transformations
\be
\d _ d \phi = - f ^ \m \p _ \m \phi \ .
\ee
As a result, equation~\eqref{conf_S} can be rewritten as
\be
0 = \int d ^ n x \, \f {\nabla f} {n} \l ( 2 \, \f {\d S} {\d g _ {\m \n}} g _ {\m \n} - \D \, \f {\d S} {\d \phi} \phi \r ) \ ,
\ee
where we used the fact that the $\d _ d \phi$ transformations can be compensated for by the corresponding transformations of the metric (provided the theory is diffeomorphism invariant). It is clear that Weyl invariance implies conformal invariance, but not the other way around, since $\nabla f $ is not an arbitrary function of coordinates.

\section{Examples}

Let us present another way to understand why Weyl invariance  necessarily implies conformal invariance in flat space-time. The corresponding conformal Killing equations now read
\be
\p _ \m \eps _ \n + \p _ \n \eps _ \m = \f {2} {n} \eta _ {\m \n} \p_\lambda \eps^\lambda \ ,
\ee
with $\eta_{\m\n}=\text{diag}\l(1,-1,\ldots\r)$, the Minkowski metric, and $\e ^ \m$ being the flat space-time analog of~$f ^ \m$. This set of equations has the following $(n+1)(n+2)/2$ parametric solution for $n \neq 2$
\be
\eps ^  \m = a ^ \m + \omega ^ \m _ {~\n} x ^ \n + c x ^ \m + 2 (b\cdot x) x ^ \m - x ^ 2 b ^ \m \ .
\label{Killings_flat}
\ee
Here $a ^ \m$, $\omega _ {\m\n} = -\omega _ {\n \m}$, $c$ and $b ^ \m$ are constants corresponding to translations, Lorentz transformations, dilatations and special conformal transformations (SCT) respectively. In two dimensions, $\e ^ \m$ is given by an arbitrary generalized harmonic function.\footnote{An example of the integrated version of the equation~\eqref{delta_c_phi} is the transformation of a scalar field under the SCT which is given by
\be
\phi ' (x') = (1 - 2\, b\cdot x + b ^ 2 x^2)^ {\D} \phi (x) \ , ~~~\text{with}~~~ x'^{\m} = \f {x ^ \m - b ^ \m x^2} {1 - 2\, b\cdot x + b ^ 2 x^2} \ .
\ee}

The standard procedure allows one to build the energy-momentum tensor
\be
\Theta _ {\m \n} = 2 \f {\d S} {\d g ^ {\m \n}} \Big | _ {g _ {\m \n} = \eta _ {\m \n}} \ ,
\ee
which is automatically traceless on the equations of motion, see~\eqref{conf_W}. As a result, all currents of the form $j _ \m = \Theta _ {\m \n} \eps ^ \n$, with $\e ^ \m$ given in~\eqref{Killings_flat}, are conserved.  

Conversely, if a theory is conformally invariant, then according to~\cite{Wess:conf,Polchinski:1987dy}, it is possible to write all currents corresponding to the conformal group in the following way
\be
j _ \m = T _ {\m \n} \eps ^ \n - \p \eps K _ \m + \p ^ \n \p \eps L _ {\m \n} \ ,
\label{conserved_curr}
\ee
where $T _ {\m \n}$ is the energy-momentum tensor (not necessarily traceless), $K _ \m$ is a vector and $L _ {\m \n}$ is a rank-two tensor such that\footnote{For $n=2$, there is an additional restriction 
\be
L _ {\m \n} = \eta _ {\m \n} L \ ,
\label{T_K_L_2}
\ee
with $L$ being a scalar function.
}
\be
\p _ \m T _ {\m \n} = 0\ , \ \ \  T _ {\m \n } = T _ {\n \m}\ , \ \ \ T _ {\m} ^ \m = n \, \p _ \m K ^ \m \ \ \ \text{and} \ \ \ K _ \m = \p ^ \n L _ {\n \m} \ .
\label{T_K_L_n}
\ee
These conditions allow to construct the improved (traceless) energy momentum tensor $\Theta _ {\m \n}$. 

However, it is not guaranteed that the theory can be made Weyl invariant. In what follows, we will consider several examples of conformally invariant theories which cannot be made Weyl invariant when coupled to gravity. We should mention though,  that we will not consider theories with non-linearly realized space-time symmetries, like in the case of galileons~\cite{Nicolis:2008in}. There, the reason that the conformal invariance of a certain action for the galileon does not imply Weyl invariance, is associated with the fact that this action is actually a Wess-Zumino term, see also~\cite{Goon:2012dy}.

\subsection{$\Box$}
\label{subsec:box}

For the purposes of illustration, it is instructive to begin by considering the Lagrangian of a free massless field in a one-dimensional spacetime
\be
\label{box}
\mc L =\f{\dot\phi^2}{2} \ . 
\ee
If the scaling dimension of $\phi$ is $\D=-1/2$, then the theory is invariant under the one-dimensional conformal group
\be
\d\phi=-\l(\eps \dot\phi-\f{1}{2}\phi \dot\eps\r) \ ,
\ee
where
\be
\eps= a+b\, t+\f{c}{2} \, t^2 \ ,
\ee
with $a,b$ and $c$, constants. The conserved currents associated with translations, dilatations and special conformal transformations  can be written according to~\eqref{conserved_curr} as
\be
J = \frac{\dot \phi ^ 2}{2} \eps - \f{\phi\dot\phi}{2} \dot \eps + \f{\phi^2}{4} \ddot \eps.
\ee
Clearly, this theory cannot be made Weyl invariant, for there are no geometric structures in $n=1$ one could use to account for the non-invariance of $\dot \phi ^ 2$.

\subsection{$\Box ^ 2$}
\label{subsec:box2}

Let us now consider the theory given by the following Lagrangian
\be
\mc L _ {\Box^2} = \f {1} {2} ( \Box \phi ) ^2 \ ,
\label{box2}
\ee
with $\Box=\eta^{\m\n}\p_\m \p_\n$ the D'Alembertian. Using the flat space-time analog of formula~\eqref{delta_c_phi} with 
$\D = n/2 - 2$, it is straightforward to check that in $n \neq 2$, the variation of this Lagrangian is given by a total derivative
\be
\d \mc L _ {\B ^ 2} = - \p^ \m \l [ \eps _ \m \mc L _ {\Box^2} - \f {2} {n} \p ^ \n \p \e \l ( \p _ \m \phi \p _ \n \phi - \f {1} {2} \eta _ {\m \n} ( \p \phi ) ^ 2 \r ) \r ] \ .
\ee
In this case, using the following definitions
\be
\begin{aligned}
T _ {\m \n} & = \eta _ {\m \n} \l( \p _ \lambda \phi \p ^ \lambda \B \phi  + \f {1} {2}  \l ( \B \phi \r ) ^ 2\r ) - \p _ \m \B  \phi \p _ \n \phi - \p _ \n \B \phi \p _ \m \phi \ , \\
K _ \m & = \f {1} {2} \B \phi \p _ \m \phi + \f {\D} {n}  \phi \p _ \m \B  \phi \ , \\
L _ {\m \n} & = \f {1} {n} \l ( 2 \p _ \n \phi \p _ \m \phi - \eta _ {\m \n} \l ( \p \phi \r ) ^ 2 + \D \eta _ {\m \n} \phi \B  \phi \r ) \ ,
\label{K_L_b2}
\end{aligned}
\ee
it is straightforward to check that the relations presented in~\eqref{T_K_L_n} are satisfied. Therefore, the system is indeed conformally invariant for $n \neq 2$.

In order to couple the theory~\eqref{box2} to gravity in a Weyl invariant fashion, we write down the most general action with four derivatives and demand that it be invariant under Weyl rescalings.\footnote{For an alternative see~\cite{Karananas:2015eha}.} As a result, we get (neglecting a term proportional to Weyl tensor squared)
\be
S _ {\Box ^ 2} = \int d ^ n x \sqrt {g} \phi \, \mc Q_4 (g) \, \phi \ ,
\ee
with
\be
\mc Q_4 (g) = \nabla ^ 4 + \nabla ^ \m \l[\l ( \f {4} {n-2} S _ {\m \n} - S g _ {\m \n} \r ) \nabla ^ \n\r] - \f {n-4} {2 (n-2)} \nabla ^ 2 S - 
\f {n-4} {(n-2)^2} S _ {\m \n} S ^ {\m \n} +  \f {n (n-4)} {4 (n-2)^2} S ^ 2 \ ,
\ee
being the Paneitz operator~\cite{Paneitz:1983_2008}, which is the Weyl covariant generalization of $\Box ^ 2$.\footnote{In a four dimensional  space-time, the Paneitz operator is also known as Paneitz-Riegert operator and it was constructed by different authors~\cite{Fradkin:1981jc,*Fradkin:1981iu,*Fradkin:1982xc,Riegert:1984kt}.} We observe that the coefficients in front of the Schouten tensor
\be
S _ {\m \n} = R _ {\m \n} - \f {1} {2 (n-1)} g _ {\m \n} R \ ,
\label{Schouten}
\ee
diverge when $n=2$. At the same time, the Schouten tensor itself vanishes due to the following relation between Ricci curvatures in two dimensions
\be
\label{ricc_2d}
R_{\m\n}=\f R 2 g_{\m\n} \ .
\ee
Therefore, the limit $n \to 2$ has to be examined separately. The most general anzatz for the operator $Q (g)$ in two dimensions has the following form
\be
\mc Q_4(g) =\nabla^4 +c_1\nabla^\m\l( R\nabla_\m\r)+c_2 \nabla^2 R +c_3 R^2 \ ,
\ee
with $c_1, c_2$ and $c_3$ constants. A straightforward calculation shows the Weyl variation of $\nabla^4$ will produce terms that cannot be cancelled by the variation of $R$-dependent terms, for example $\l ( \nabla^\m \nabla _ \n \s \r ) \nabla _ \m \nabla _ \n$. Therefore, for $n=2$ there is no Weyl covariant generalization of the fourth-order differential operator. Hence, in this case, the system~\eqref{box2}  cannot be coupled to gravity in a Weyl invariant way, although this does not come as a surprise, for as it is clear from~\eqref{K_L_b2}, the condition~\eqref{T_K_L_2} is not satisfied. One can say that the system at hand in a two dimensional space-time, is only invariant under global conformal transformations, which correspond to the six dimensional sub-algebra of the Virasoro algebra.

\subsection{$\Box ^ 3$}

The fact that it is impossible to construct a Weyl invariant action for the system~\eqref{box2} in two dimensions, is a particular case of a more general result~\cite{GJMS,GJMS_2,GJMS_3}, see also~\cite{Nakayama:2013is}. This states that for even number of dimensions, there exist Weyl invariant generalizations of $\Box ^ k$ only for 
$k \leq \f {n} {2}$. Therefore, considering a theory with six derivatives in a four dimensional space-time
\be
\mc L _ {\B ^ 3}= \f {1} {2} \l ( \p _ \m{\B} \phi \r ) ^ 2 \ ,
\label{box3}
\ee
one is sure that it cannot be made Weyl invariant. This can be immediately seen by inspecting the Weyl covariant analog of the operator~\eqref{box3}.\footnote{Explicit expressions for the operator have been obtained in~\cite{Branson:1985,Osborn:2015rna}.} It contains terms proportional to
\be
\f{1}{(n-2)(n-4)}B_{\m\n}S^{\m\n} \ ,~~~\f{1}{n-4}\nabla^\m\l(B_{\m\n}\nabla^\n\r) \ ,
\label{singular_terms}
\ee
thus it does not exist in $n=2$ and $n=4$ dimensions for a non-zero Bach tensor $B_{\m\n}$ 
\be
B_{\m\n}=W_{\m\rho\n\s}S^{\rho\s} +\nabla^\rho \nabla_\m S_{\n\rho}-\nabla^2 S_{\m\n}\ ,
\label{Bach_tensor}
\ee
with $W_{\m\rho\n\s}$ being the Weyl tensor.

However, straightforward computations reveal -- taking into account that the scaling dimension of the field in this case is equal to $\D = n/2 - 3$ -- that the conformal variation of the Lagrangian~\eqref{box3}  is also a total derivative
\be
\d \mc L _ {\B ^ 3} = - \p ^ \m \l [ \eps _ \m \mc L _ {\Box^3} - 
\f {1} {n} \p ^ \n \p \e \l ( 4 \p _ \m \p _ \n \phi \B \phi  - \f {1} {2} ( \B \phi ) ^ 2 \l  ( \f {n} {2} + 3 \r ) \eta _ {\m \n} \r ) \r ] \ .
\ee
Moreover, one can build the energy-momentum tensor
\be
\begin{aligned}
T _ {\m \n} &= \Box^2 \phi \, \p _ \m \p _ \n \phi - \l ( \p _ \m \phi \, \p _ \n \B^2 \phi + \p _ \n \phi \, \p _ \m \B ^ 2 \phi \r ) \\
&+ \p ^\lambda \phi \, \p _ \m \p _ \n \p _ \lambda \B \phi + \B \phi \, \p _ \m \p _ \n \B \phi + \p ^ \lambda \B \phi \, \p _ \m \p _ \n \p _ \lambda \phi \\
&- \p _ \m \B \phi \, \p _ \n \B \phi - \eta _ {\m \n} \l [ \f {1} {2} \l( \p _ \lambda \B \phi \r ) ^ 2 + \p ^\lambda \p ^ \s \phi \, \p _ \lambda \p _ \s \B \phi \r ] \ ,
\end{aligned}
\ee
as well as the operators
\be
\begin{aligned}
K _ \m &= \a \, \p _ \m \p _ \n \phi \, \p _ \n \B  \phi - (n+\a) \p ^ \n \phi \,  \p _ \m \p _ \n \B  \phi - 
\l ( \f {n} {2} + \a \r ) \p _ \m \B \phi \, \B  \phi  \\
&+ \l ( \a + \f {n} {2} + 2 \r ) \p _ \m \phi \, \B ^ 2 \phi + \l ( \f {n} {2} - 3 \r ) \phi \, \p _ \m \B ^ 2 \phi \ , 
\end{aligned}
\ee
and
\be
\begin{aligned}
L _ {\m \n} &=& \l ( \a - \f {n-10} {4} \r ) \p _ \m \phi \, \p _ \n \B  \phi - \l ( \a + \f {3 n - 10 } {4} \r ) \p _ \n \phi \, \p _ \m \B  \phi  \nn \\
&+& \f {n-10} {4} \p _ \m \p _ \n \phi \, \B  \phi - \f {n+10} {4} \phi \, \p _ \m \p _ \n \B  \phi 
+ \f {3n - 2} {4} \eta _ {\m \n} \phi \, \B ^ 2 \phi \ .
\end{aligned}
\ee
The above satisfy~\eqref{T_K_L_n} for arbitrary values of the constant $\a$,  therefore, the theory is conformal in flat space-time. Notice, though, that for $L_{\m\n}$ to be symmetric, we have to set $\a=-n/4$.

\subsection{Curved space-time}

In order to further expose the difference between the concepts of Weyl and conformal symmetries we consider the curved space-time counterpart of $\B^3$. It is obvious that the sixth-order Weyl covariant operator for $n\neq 2$ and $n \neq 4$ is also conformally invariant for an arbitrary metric. It may happen though that there are no conformal Killings for a specific background to start with. To guarantee that the conformal group is not empty, we stick to Einstein manifolds only, for which
\be
\label{ein-man-1}
R_{\m\n}=\f {R}{n} g_{\m\n} \ . 
\ee
It is easy to check that the Bach tensor~\eqref{Bach_tensor} in this case vanishes identically.\footnote{To make this point clear, we proceed as follows. For Einstein manifolds, the Schouten tensor~\eqref{Schouten} becomes
$$
S_{\m\n}=\f{n-2}{2n(n-1)} R g_{\m\n} \ .
$$
Upon plugging the above into the definition of the Bach tensor~\eqref{Bach_tensor} and recalling that the Weyl tensor is traceless in all of its indices, we find that 
$$
B_{\m\n}=\f{n-2}{2n(n-1)}\l(\nabla_\m\nabla_\n R-g_{\m\n}\nabla^2 R \r) \ ,
$$
which is zero for all $n$. This follows trivially from the (contracted) Bianchi identities, which yield that the scalar curvature $R$ is constant (for $n\neq 2$).} Therefore, the dangerous terms~\eqref{singular_terms} disappear, thus the limit 
$n \to 4$ of the conformally invariant curved space analog of $\Box ^3$, can be safely considered. In doing so, one obtains a conformally invariant operator with leading term $\nabla ^ 6$.

To illustrate the procedure in more detail, we consider the Paneitz operator, which for Einstein manifolds becomes regular at $n=2$. It is straightforward to check using the relation
\be
\label{ein-man-conf_K}
\nabla_\m \nabla_\n \nabla f + \f{1}{2n(n-1)} g_{\m\n}\l(n\, f^\s \nabla_\s R+2\,R\nabla f\r) = 0 \ ,
\ee
following from the conformal Killing equations for $n\neq 2$, that the corresponding action
\be
\begin{aligned}
\label{ein-man-3}
S=\int d^n x\sqrt{g}\Bigg[\nabla^2\phi\nabla^2\phi&-\f{4-n(n-2)}{2n(n-1)}R\l(\nabla\phi\r)^2  -\f{n-4}{4(n-1)}\phi^2\nabla^2 R \\
&+\f{(n-2)(n+2)(n-4)}{16n(n-1)^2}\phi^2R^2\Bigg]\ ,
\end{aligned}
\ee
is invariant under the ($n \neq 2$) conformal transformations, as it should. The limit $n \to 2$ in turn is regular
\be
\begin{aligned}
S=\int d^n x\sqrt{g}\Bigg[\nabla^2\phi\nabla^2\phi&-R\l(\nabla\phi\r)^2  +\f{1}{2}\phi^2\nabla^2 R \Bigg],
\end{aligned}
\ee
and is invariant under global conformal transformations.
The reason it is not invariant under the full conformal group is that the relation~\eqref{ein-man-conf_K} does not follow automatically for two dimensional theories. Rather, it has to be imposed by hand, reducing the conformal group to its subgroup of global transformations. Clearly this is a peculiarity of two dimensions.

\section{Generalization}

The examples we considered clearly show that not any conformally invariant (both in flat and curved space-time) theory can be made Weyl invariant. In fact, there is a whole class of theories not allowing Weyl invariant generalizations. Indeed, as it was mentioned before, according
to~\cite{GJMS,GJMS_2,GJMS_3}, the Weyl covariant analogs of $\Box^k$ exist unless the number of space-time dimensions $n$ is even and less than $k/2$. The impossibility to construct the corresponding operators in even number of dimensions manifests itself through the presence of terms singular at $n = 2,4,6,\ldots$ However, it seems plausible that similar to the situation described in the previous section those terms vanish (or at least become regular) once the geometry is restricted to that of Einstein spaces. As a result, the corresponding limit $n \to 4,6,\ldots$ exists and is invariant under conformal transformations (or only global conformal transformations for $n \to 2$).

Since flat spaces are a particular case of Einstein ones, according to the above argument, the theories whose dynamics is described by the Lagrangian in flat space-time
\be
\mc L _ {\Box ^ {k}} = \f {1} {2} \phi \Box ^ k \phi \ 
\label{conf_k}
\ee
are conformal (for $n \neq 2$). We can convince ourselves that this is the case by considering the variation of the Lagrangian with respect to conformal transformations. For $k = 2 m$ and $k = 2 m + 1$, the Lagrangian can be rewritten as
\be
\mc L _ {\Box ^ {2m}} = \f {1} {2} ( \Box ^ m \phi ) ^ 2 ~~ \text {and} ~~ \mc L _ {\Box ^ {2m+1}} = \f {1} {2} ( \p_\m \Box ^ m \phi ) ^ 2,
\ee
while the variations are respectively given by
\be
\d _ c \mc L _ {\Box ^ {2 m}} = - \p _ \m \l \{ \e ^ \m \mc L _ {\Box ^ {2 m}} - \f {2 m^2} {n} \, \p _ \n \p \e \, \l [ \p ^ \m \Box ^ {m-1} \phi \p ^ \n \Box ^ {m-1} \phi - \f {1} {2} \eta ^ {\m \n} \l ( \p \Box ^ {m-1} \phi \r ) ^ 2 \r ] \r \} \ ,
\label{conf_2m}
\ee
and
\be
\label{conf_2m1}
\begin{aligned}
\d _ c \mc L _ {\Box ^ {2 m + 1}} = - \p _ \m \Bigg \{ \e ^ \m \mc L _ {\Box ^ {2 m + 1}} - \f {1} {n} \, \p _ \n \p \e & \bigg [ {2 m (m+1)} \p ^ \m \p ^ \n \Box ^ {m-1} \phi \Box ^ {m} \phi  \\
& - \f {1} {2} \eta ^ {\m \n} \l ( \f {n} {2} - 1 + {2 m (m+1)} \r ) \l ( \Box ^ {m} \phi \r ) ^ 2 \bigg ] \Bigg \} \ . 
\end{aligned}
\ee
At the same time, according to~\cite{GJMS_3}, the Lagrangian~\eqref{conf_k} cannot be made Weyl invariant in an even number of dimensions if $n < 2 k$.

Similarly, it can be proven that for manifolds with vanishing Ricci tensor, the theories given by the Lagrangian
\be
\mc L _ {\nabla ^ {2k}} = \f {1} {2} \phi \nabla ^ {2k} \phi \ ,
\ee
are also conformally invariant.

\section*{Acknowledgements}

We would like to thank R.~Rattazzi and M.~Shaposhnikov for useful discussions. The work of G.K.K. and A.M. is supported by the Swiss National Science Foundation.

\bibliographystyle{utphys}
\bibliography{higher_derivatives}{}

\providecommand{\href}[2]{#2}\begingroup\raggedright\begin{thebibliography}{10}

\bibitem{Wess:conf}
J.~Wess, ``{Conformal Invariance in Quantum Field Theory},'' {\em Nuovo Cim}
  {\bfseries 18} (1960) 1086.

\bibitem{Polchinski:1987dy}
J.~Polchinski, ``{Scale and Conformal Invariance in Quantum Field Theory},''
\href{http://dx.doi.org/10.1016/0550-3213(88)90179-4}{{\em Nucl. Phys.}
  {\bfseries B303} (1988) 226}.

\bibitem{Nicolis:2008in}
A.~Nicolis, R.~Rattazzi, and E.~Trincherini, ``{The Galileon as a local
  modification of gravity},''
  \href{http://dx.doi.org/10.1103/PhysRevD.79.064036}{{\em Phys. Rev.}
  {\bfseries D79} (2009) 064036},
\href{http://arxiv.org/abs/0811.2197}{{\ttfamily arXiv:0811.2197 [hep-th]}}.

\bibitem{Goon:2012dy}
G.~Goon, K.~Hinterbichler, A.~Joyce, and M.~Trodden, ``{Galileons as
  Wess-Zumino Terms},'' \href{http://dx.doi.org/10.1007/JHEP06(2012)004}{{\em
  JHEP} {\bfseries 06} (2012) 004},
\href{http://arxiv.org/abs/1203.3191}{{\ttfamily arXiv:1203.3191 [hep-th]}}.

\bibitem{Karananas:2015eha}
G.~K. Karananas and A.~Monin, ``{Weyl and Ricci gauging from the coset
  construction},'' \href{http://dx.doi.org/10.1103/PhysRevD.93.064013}{{\em
  Phys. Rev.} {\bfseries D93} no.~6, (2016) 064013},
\href{http://arxiv.org/abs/1510.07589}{{\ttfamily arXiv:1510.07589 [hep-th]}}.

\bibitem{Paneitz:1983_2008}
S.~M. Paneitz, ``{A Quartic Conformally Covariant Differential Operator for
  Arbitrary Pseudo-Riemannian Manifolds (Summary)},'' {\em SIGMA, Symmetry,
  Integrability and Geometry: Methods and Applications} {\bfseries 4} (2008) 3.

\bibitem{Fradkin:1981jc}
E.~S. Fradkin and A.~A. Tseytlin, ``{One Loop Beta Function in Conformal
  Supergravities},''
\href{http://dx.doi.org/10.1016/0550-3213(82)90481-3}{{\em Nucl. Phys.}
  {\bfseries B203} (1982) 157}.

\bibitem{Fradkin:1981iu}
E.~S. Fradkin and A.~A. Tseytlin, ``{Renormalizable asymptotically free quantum
  theory of gravity},''
\href{http://dx.doi.org/10.1016/0550-3213(82)90444-8}{{\em Nucl. Phys.}
  {\bfseries B201} (1982) 469--491}.

\bibitem{Fradkin:1982xc}
E.~S. Fradkin and A.~A. Tseytlin, ``{Asymptotic freedom in extended conformal
  supergravities},''
\href{http://dx.doi.org/10.1016/0370-2693(82)91018-8}{{\em Phys. Lett.}
  {\bfseries B110} (1982) 117--122}.

\bibitem{Riegert:1984kt}
R.~J. Riegert, ``{A Nonlocal Action for the Trace Anomaly},''
\href{http://dx.doi.org/10.1016/0370-2693(84)90983-3}{{\em Phys. Lett.}
  {\bfseries B134} (1984) 56--60}.

\bibitem{GJMS}
C.~R. Graham, R.~Jenne, L.~J. Mason, and G.~A.~J. Sparling, ``{Conformally
  Invariant Powers of the Laplacian I: Existence},''
  \href{http://dx.doi.org/10.1112/jlms/s2-46.3.557}{{\em J. London Math. Soc.}
  {\bfseries s2-46 (3)} (1992) 557--565}.

\bibitem{GJMS_2}
C.~R. Graham, ``{Conformally Invariant Powers of the Laplacian II:
  Nonexistence},'' \href{http://dx.doi.org/10.1112/jlms/s2-46.3.566}{{\em J.
  London Math. Soc.} {\bfseries s2-46 (3)} (1992) 566--576}.

\bibitem{GJMS_3}
A.~R. Gover and K.~Hirachi, ``{Conformally Invariant Powers of the Laplacian --
  A Complete Nonexistence Theorem},'' {\em Journa of the American Mathematical
  Society} {\bfseries 17} no.~2, (2004) 389--405.

\bibitem{Nakayama:2013is}
Y.~Nakayama, ``{Scale invariance vs conformal invariance},''
  \href{http://dx.doi.org/10.1016/j.physrep.2014.12.003}{{\em Phys. Rept.}
  {\bfseries 569} (2015) 1--93},
\href{http://arxiv.org/abs/1302.0884}{{\ttfamily arXiv:1302.0884 [hep-th]}}.

\bibitem{Branson:1985}
T.~Branson, ``{Differential Operators Canonically Associated to a Conformal
  Structure},'' {\em Math.Scand.} {\bfseries 57} (1985) 293.

\bibitem{Osborn:2015rna}
H.~Osborn and A.~Stergiou, ``{Structures on the Conformal Manifold in Six
  Dimensional Theories},''
  \href{http://dx.doi.org/10.1007/JHEP04(2015)157}{{\em JHEP} {\bfseries 04}
  (2015) 157},
\href{http://arxiv.org/abs/1501.01308}{{\ttfamily arXiv:1501.01308 [hep-th]}}.

\end{thebibliography}\endgroup

\end{document}